\documentclass[prd,twocolumn,showpacs,amsmath,amssymb]{revtex4-1}
\usepackage{dcolumn}
\usepackage{bm}

\usepackage[utf8]{inputenc}
\usepackage[T1]{fontenc}
\usepackage{chngpage}
\usepackage{float}
\usepackage{fancyvrb}
\usepackage{hyperref}
\usepackage[pdftex]{graphicx}
\usepackage{epstopdf}
\usepackage{physics}
\usepackage{listings}
\usepackage{color}
\usepackage{caption}
\usepackage{subcaption}
\usepackage{breqn}
\usepackage{framed}
\usepackage{tocloft}
\usepackage{wasysym}
\usepackage{overpic}

\newcommand{\asymm}{\mathcal{A}_\varphi}

\newcommand{\sintwophi}{\sin 2 \varphi}

\begin{document}
\normalsize
\parskip=5pt

\title{\boldmath Measurement of the Branching Fraction of and Search for a CP-Violating Asymmetry in $\eta' \to \pi^+ \pi^- e^+ e^-$ at BESIII}
\author{\small
M.~Ablikim$^{1}$, M.~N.~Achasov$^{10,c}$, P.~Adlarson$^{64}$, S. ~Ahmed$^{15}$, M.~Albrecht$^{4}$, A.~Amoroso$^{63A,63C}$, Q.~An$^{60,47}$, X.~H.~Bai$^{54}$, Y.~Bai$^{46}$, O.~Bakina$^{29}$
% ...
\\
\vspace{0.2cm}
(BESIII Collaboration)\\
\vspace{0.2cm} {\it
$^{1}$ Institute of High Energy Physics, Beijing 100049, People's Republic of China\\
% ...
$^{4}$ Bochum Ruhr-University, D-44780 Bochum, Germany\\
% ...
$^{10}$ G.I. Budker Institute of Nuclear Physics SB RAS (BINP), Novosibirsk 630090, Russia\\
% ...
$^{15}$ Helmholtz Institute Mainz, Johann-Joachim-Becher-Weg 45, D-55099 Mainz, Germany\\
% ...
$^{29}$ Joint Institute for Nuclear Research, 141980 Dubna, Moscow region, Russia\\
% ...
$^{46}$ Southeast University, Nanjing 211100, People's Republic of China\\
$^{47}$ State Key Laboratory of Particle Detection and Electronics, Beijing 100049, Hefei 230026, People's Republic of China\\
% ...
$^{54}$ University of Jinan, Jinan 250022, People's Republic of China\\
% ...
$^{60}$ University of Science and Technology of China, Hefei 230026, People's Republic of China\\
% ...
$^{63}$ University of Turin and INFN, (A)University of Turin, I-10125, Turin, Italy; (B)University of Eastern Piedmont, I-15121, Alessandria, Italy; (C)INFN, I-10125, Turin, Italy\\
$^{64}$ Uppsala University, Box 516, SE-75120 Uppsala, Sweden\\
% ...
$^{c}$ Also at the Novosibirsk State University, Novosibirsk, 630090, Russia\\
% ...
}
}
\date{\today}

\begin{abstract}
The rare decay $\eta' \to \pi^+ \pi^- e^+ e^-$ is studied using a sample of $1.3 \times 10^9$ $J/\psi$ events collected with the BESIII detector at BEPCII in 2009 and 2012. The branching fraction is measured with improved precision to be $\left( 2.42\pm0.05_{stat.}\pm0.08_{syst.} \right) \times 10^{-3}$. Due to the inclusion of new data, this result supersedes the last \mbox{BESIII} result on this branching fraction. In addition, the CP-violating asymmetry in the angle between the decay planes of the $\pi^+\pi^-$-pair and the $e^+e^-$-pair is investigated.  A measurable value would indicate physics beyond the standard model; the result is $\mathcal{A}_{CP} = \left(2.9\pm3.7_{stat.}\pm1.1_{syst.} \right)\%$,
which is consistent with the standard model expectation of no CP-violation. The precision is comparable to the asymmetry measurement 
in the $K^0_L\to\pi^+\pi^-e^+e^-$ decay where the observed $(14\pm2)\%$ effect is driven by a standard model mechanism.
\end{abstract}


\maketitle

\section{Introduction}
At the precision/intensity frontier of particle physics, rare decays of light mesons such as the $\eta'$ are valuable laboratories for the study of a number of interesting phenomena.  They offer a way of testing the effective field theories that describe quantum chromodynamics at low energies.  
In addition, they provide an excellent window for physics beyond the standard model (SM), since small SM rates make possible new contributions easier to detect.  

The decay $\eta' \to \pi^+ \pi^-\gamma^* \to \pi^+ \pi^- e^+ e^-$, can be described both by vector meson dominance (VMD) models~\cite{SAKURAI19601, Petri:2010ea} and by unitary chiral perturbation theory~\cite{Borasoy:2007dw}. Precise experimental data is needed to guide phenomenological descriptions of the reaction. The process $\eta'\to\pi^+\pi^- e^+ e^-$ is expected to proceed through an intermediate virtual photon, with dynamics closely related to the decay $\eta' \to \pi^+ \pi^- \gamma$ which has been studied recently by the BESIII experiment~\cite{Ablikim:2017fll}. As in the $\eta' \to \pi^+ \pi^- \gamma$ decay, $\eta' \to \pi^+ \pi^- e^+ e^-$ is expected to exhibit a contribution from the box anomaly of the Wess-Zumino-Witten Lagrangian~\cite{Witten:1983tw} and a dominant $\rho\to\pi^+\pi^-$ contribution.

\begin{figure}
\centering
\includegraphics[scale=0.4]{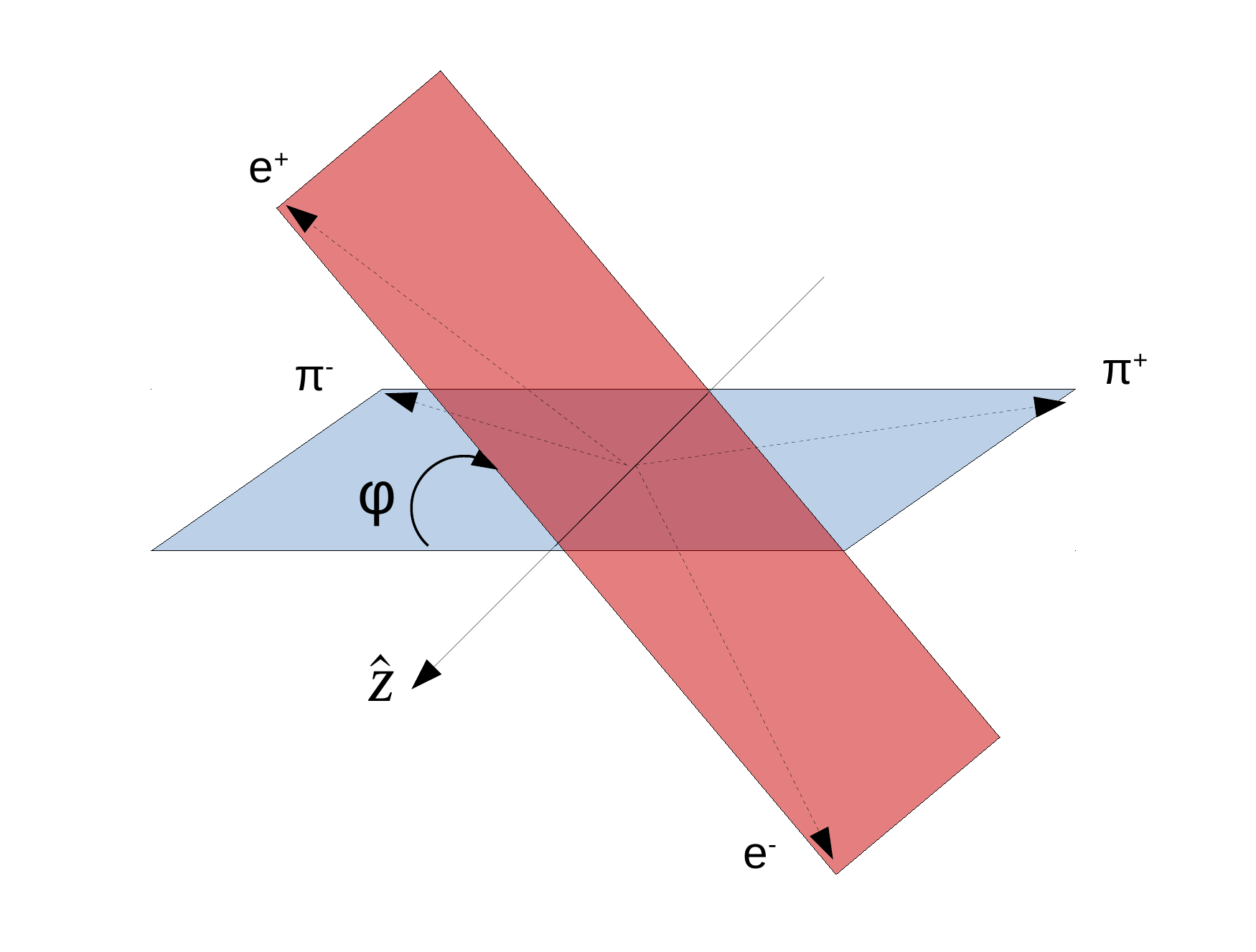}
\caption{Illustration of the decay plane angle $\varphi$. $\hat{z}$ is a unit vector along the intersection of the two decay planes. $\varphi$ can be unambiguously determined from Eqs.~\ref{eq:Sin2phiBegin} to~\ref{eq:Sin2phiEnd}.}
\label{fig:AsymmIll}
\end{figure}

A new aspect of $\eta' \to \pi^+ \pi^- e^+ e^-$ compared to  $\eta' \to \pi^+ \pi^- \gamma$ is that in analogy with the decay $K^0_L \to \pi^+ \pi^- \gamma^* \, \to \pi^+ \pi^- e^+ e^-$ studied in Refs.~\cite{PhysRevLett.84.408, IconomidouFayard:2001ww}, it allows for a test of CP-violation due to the interference between the dominating, CP-conserving, magnetic transition, and a possible CP-violating electric dipole type transition~\cite{Geng:2002ua, Gao:2002gq,Gan:2020aco}. Such an interference term is proportional to $\sintwophi$, where $\varphi$ is the angle between the decay planes of the $e^+e^-$-pair and the $\pi^+ \pi^-$-pair in the reference frame of the $\eta'$-meson as shown in Fig.~\ref{fig:AsymmIll} and unambiguously defined by Eqs~\ref{eq:Sin2phiBegin} to~\ref{eq:Sin2phiEnd}. Hence, an electric dipole transition will manifest itself as an asymmetry $\mathcal{A}_{CP}$ of the $\sintwophi$ distribution~\cite{Gao:2002gq,Petri:2010ea}
\begin{align} 
    \begin{split}
 \mathcal{A}_{CP} \quad &= \quad <\text{sgn}( \sin2\varphi)> \\
 \quad &=\quad \frac{1}{ \Gamma} \int_0^{2 \pi} \frac{d \Gamma}{d \varphi} \,  \text{sgn}( \sin 2\varphi) \, d\varphi\ ,
 \end{split}
\end{align}
where $\Gamma$ is the total decay width, sgn is the sign function, and $d\Gamma/d\varphi$ is the partial decay width for $\eta' \to \pi^+ \pi^- e^+ e^-$.
Experimentally, this quantity can be extracted as:
\begin{align}
\mathcal{A}_\varphi = \frac{N(\sin2\varphi >0) -N(\sin2\varphi<0)}{N(\sin 2\varphi >0) + N(\sin2\varphi <0)}\ ,
\label{eq:AsymmExp}
\end{align}
where $N(x)$ is the acceptance-corrected number of events in the corresponding angular region.

The first experimental evidence of the decay $\eta' \to \pi^+ \pi^- e^+ e^-$ was obtained by the CLEO experiment~\cite{Naik:2008aa}. The branching fraction was determined to be $\left(2.5^{+1.2}_{-0.9}\pm 0.5 \right) \times 10^{-3}$ from $7.9^{+3.9}_{-2.7}$ events. The most precise measurement of the branching fraction to date was made by the \mbox{BESIII} experiment using 225 million $J/\psi$ events collected in 2009~\cite{Ablikim:2013wfg}. The branching fraction was determined to be $(2.11\pm 0.12 (stat.) \pm 0.15 (syst.)) \times 10^{-3}$ from $429 \pm 24$ events. 

The asymmetry $\mathcal{A}_\varphi$ has been measured for the decay $\eta \to \pi^+ \pi^- e^+e^-$ by the WASA-at-COSY~\cite{Adlarson:2015zta} and KLOE~\cite{Ambrosino2009283} experiments, and was found to be consistent with zero in both cases.  The asymmetry has never been measured in the decay $\eta' \to \pi^+ \pi^- e^+ e^-$. This work aims to reduce the statistical uncertainty on the branching fraction by the inclusion of an additional 1.085 billion $J/\psi$ events collected by the \mbox{BESIII} experiment in 2012~\cite{Ablikim:2016fal, Ablikim:2019hff} and perform the first measurement of the asymmetry $\mathcal{A}_\varphi$.

\section{Detector and Data Samples}

In this work, $1.31 \times 10^9$ $J/\psi$ meson decays ~\cite{Ablikim:2016fal, Ablikim:2019hff} collected by BESIII in 2009 and 2012 are analyzed. Through the radiative decay $J/\psi \to \gamma \eta' $, this yields a sample of about $6.7$ million $\eta'$ events.

The BESIII detector is a magnetic
spectrometer~\cite{Ablikim:2009aa} collecting data at the Beijing Electron
Positron Collider II (BEPCII)~\cite{Yu:2016cof}. 
It covers 93\% of the solid angle, and consists of a helium-based
 multilayer drift chamber (MDC), a plastic scintillator time-of-flight
system (TOF), and a CsI(Tl) electromagnetic calorimeter (EMC),
which are all enclosed in a superconducting solenoidal magnet
providing a 1.0~T (0.9~T in 2012) magnetic field. The solenoid is supported by an
octagonal flux-return yoke with resistive plate counter muon
identifier modules interleaved with steel. The
charged-particle momentum resolution at $1~{\rm GeV}/c$ is
$0.5\%$, and the $dE/dx$ resolution is $6\%$ for the electrons
from Bhabha scattering. The EMC measures photon energies with a
resolution of $2.5\%$ ($5\%$) at $1$~GeV in the barrel (end cap)
region. The time resolution of the TOF barrel part is 68~ps, while
that of the end cap part is 110~ps. 

For determination of detection efficiencies and estimation of background, Monte Carlo (MC) samples are produced with a {\sc geant4}-based~\cite{geant4} detector simulation package BOOST~\cite{Deng:371}, which
includes the geometry and response of the BESIII detector. The beam
energy spread and initial state radiation (ISR) in the $e^+e^-$
annihilations is modelled with the generator {\sc
kkmc}~\cite{ref:kkmc}. For qualitative studies of the background contributions, an inclusive MC sample is used. For this sample, the $J/\psi$ resonance as well as continuum processes are produced by {\sc
kkmc}~\cite{ref:kkmc}. Known decay modes are modelled with {\sc
evtgen}~\cite{ref:evtgen, evtgenweb} using branching fractions taken from the
Particle Data Group (PDG)~\cite{PDG2020}, and the remaining unknown decays
from the charmonium states are produced with {\sc
lundcharm}~\cite{ref:lundcharm}. Final state radiation (FSR)
from charged particles is incorporated with the {\sc
photos} package~\cite{photos}. The decay $J/\psi \to \gamma \eta'$ is generated with the {\sc HELAMP} model of {\sc EvtGen}~\cite{ref:MCGen}.
Specific generators based on {\sc EvtGen} have been developed for the decays $\eta' \to \pi^+ \pi^- e^+e^-$~\cite{1674-1137-36-10-002} and $\eta' \to \pi^+ \pi^- \gamma$~\cite{Ablikim:2017fll} using theoretical decay amplitudes from the VMD model~\cite{Petri:2010ea}. The MC samples referred to as {\it signal MC} throughout the text have been generated using the former generator.

\section{Analysis}
In order to minimize the impact of systematic effects due to the number of $J/\psi$ mesons, tracking and charged particle identification (PID) of pions, as well as the reconstruction of photons, the branching fraction of $\eta' \to \pi^+\pi^-e^+e^-$ is determined relative to the channel $\eta' \to \pi^+\pi^-\gamma$. This is the second most probable decay of the $\eta'$ meson, and its branching fraction is well known~\cite{Pedlar:2009aa,Ablikim:2017fll,PDG2020}.  The branching fraction for $\eta' \to \pi^+\pi^-e^+e^-$ is calculated according to
\begin{align}
\begin{split}
&{\cal B}(\eta' \rightarrow \pi^+ \pi^- e^+ e^-) = \\
&\frac{N_{\eta' \rightarrow \pi^+ \pi^- e^+ e^-} \times \varepsilon_{\eta' \rightarrow \pi^+ \pi^- \gamma} \times {\cal B}({\eta' \rightarrow \pi^+ \pi^- \gamma})}{N_{\eta' \rightarrow \pi^+ \pi^- \gamma} \times \varepsilon_{\eta' \rightarrow \pi^+ \pi^- e^+ e^-}},
\end{split}
\label{eq:BRNorm}
\end{align}
where $N_{i}$ is the event yield in channel $i$, $\varepsilon_{i}$ is the corresponding efficiency, and ${\cal B}({\eta' \rightarrow \pi^+ \pi^- \gamma})$ is the branching fraction for the decay $\eta' \rightarrow \pi^+ \pi^- \gamma$.

\subsection{Signal: $\eta' \to \pi^+ \pi^- e^+ e^-$}
The signal channel is studied through the decay chain $J/\psi \to \gamma \eta', \,  \eta' \to \pi^+ \pi^- e^+ e^-$. Each event is required to contain four charged track candidates with net charge zero, and at least one photon candidate. The MDC enables reconstruction of charged tracks within $|\cos \theta| < 0.93$, where $\theta$ is the polar angle of the track relative to the symmetry axis of the detector.
Tracks are also required to originate from a region within 10 cm of the interaction point (IP) in the longitudinal direction, and 1 cm in the transverse direction.

Photons are reconstructed from showers with a deposited energy of at least 25 MeV in the barrel EMC ($| \cos \theta | < 0.80$), or at least 50 MeV in the endcap EMC ($0.86 < |\cos \theta| < 0.92$). To ensure that these showers originate from a photon, the angle between the shower and the nearest charged track is required to be larger than $15^\circ$. Finally, photons are required to arrive within 700 ns from the event start time in order to reduce background from photons that do not originate from the same event.

For each candidate event, TOF and $dE/dx$ information is used to perform PID. Furthermore, a four-constraint (4C) kinematic fit of all final state particles to the initial $J/\psi$ four momentum is performed. For both the PID and kinematic fit,
the hypotheses $\gamma \pi^+ \pi^- e^+ e^-$, $\gamma \pi^+ \pi^- \mu^+ \mu^-$ and $\gamma \pi^+ \pi^- \pi^+ \pi^-$ are tested.
The hypotheses are evaluated with the combined $\chi^2_{4C+PID} = \chi^2_{4C} + \sum_{j=1}^{4}\chi^2_{PID}(j)$, where $\chi^2_{PID}(j)$ is the PID $\chi^2$ for track $j$.  Only those events  where the $\gamma \pi^+ \pi^- e^+ e^-$ hypothesis is best are kept. If the event contains more than one candidate photon, the one that gives the least $\chi^2_{4C}$ is selected. Finally, an upper limit on $\chi^2_{4C+PID} < 62$ is imposed. This cut-off has been optimized with respect to the figure of merit $N_{S}/\sqrt{N_{D}}$, where $N_{S}$ is the number of events in the signal MC sample, and $N_{D}$ is the number of events in data after the final selection.

\begin{figure}[tbh!]
\centering
\begin{overpic}[width=0.4\textwidth]{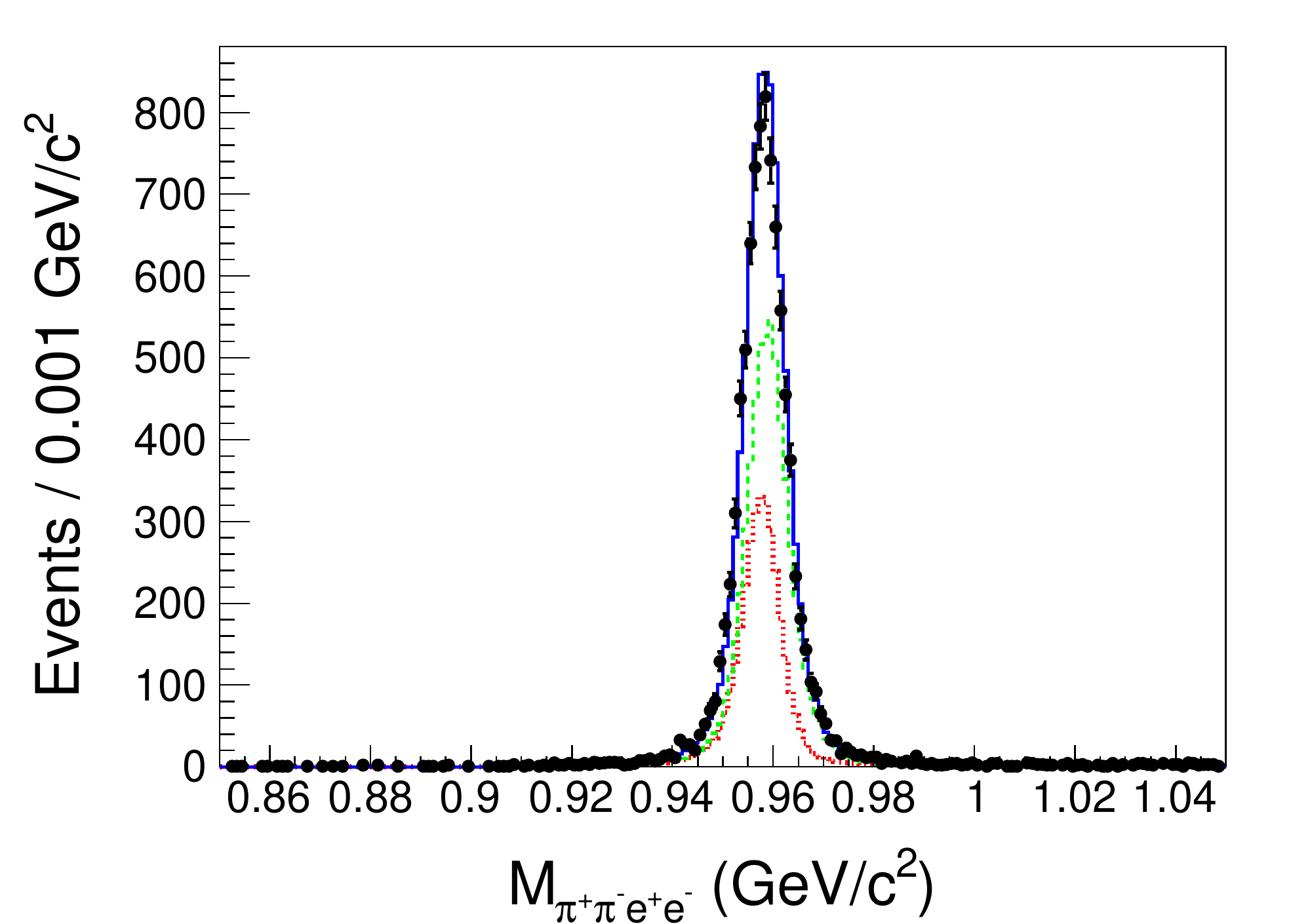}
\put(86,62){\color{red}\bf\large(a)}
\end{overpic}
\begin{overpic}[width=0.4\textwidth]{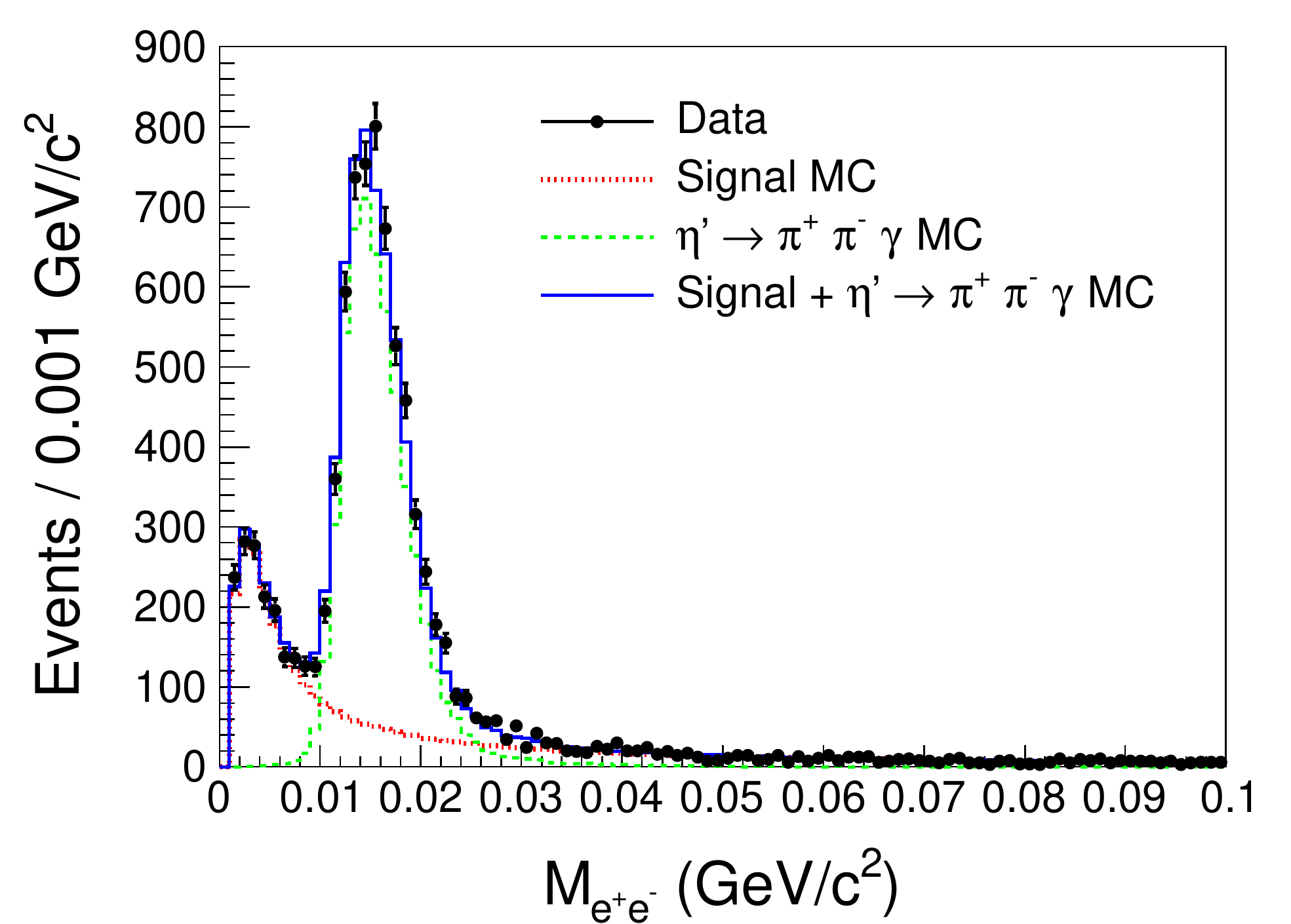}
\put(86,62){\color{red}\bf\large(b)}
\end{overpic}
\caption{Invariant mass distributions of $\pi^+\pi^-e^+e^-$ (a) and $e^+e^-$ (b) for events after the preselection. The dots with error bars correspond to data, the dotted (red) histogram represents the signal MC sample, the dashed (green) histogram represents the $\eta' \to \pi^+ \pi^- \gamma$ MC sample, and the solid (blue) histogram is the sum of the two contributions. The two contributions are normalized according to their respective branching fractions and their efficiencies as determined from the MC simulation.}
\label{fig:MeeMCData}
\end{figure}

\begin{figure}
\begin{overpic}[width=0.4\textwidth]{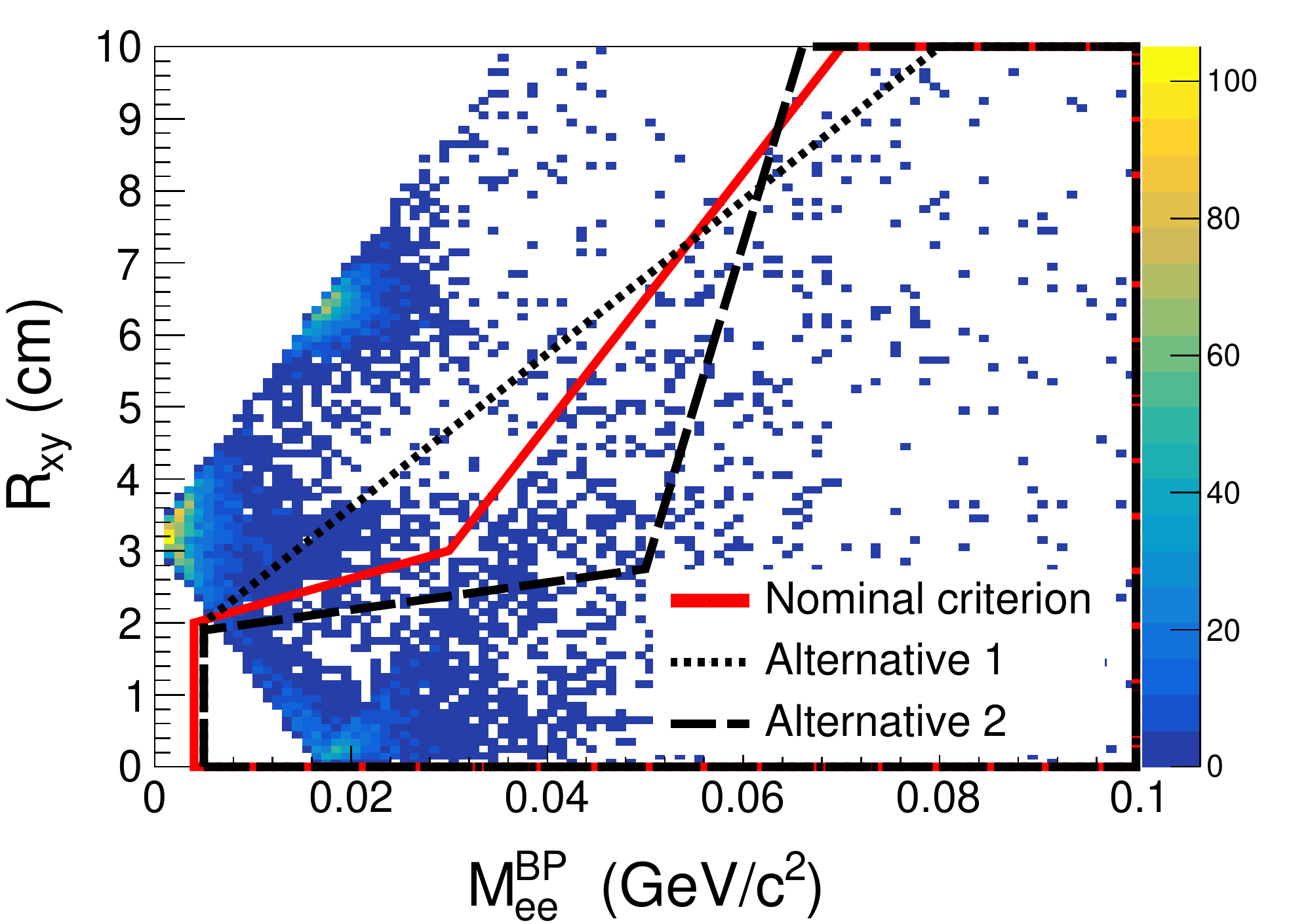}
\put(15,62){\color{red}\bf\large(a)}
\end{overpic}
\begin{overpic}[width=0.4\textwidth]{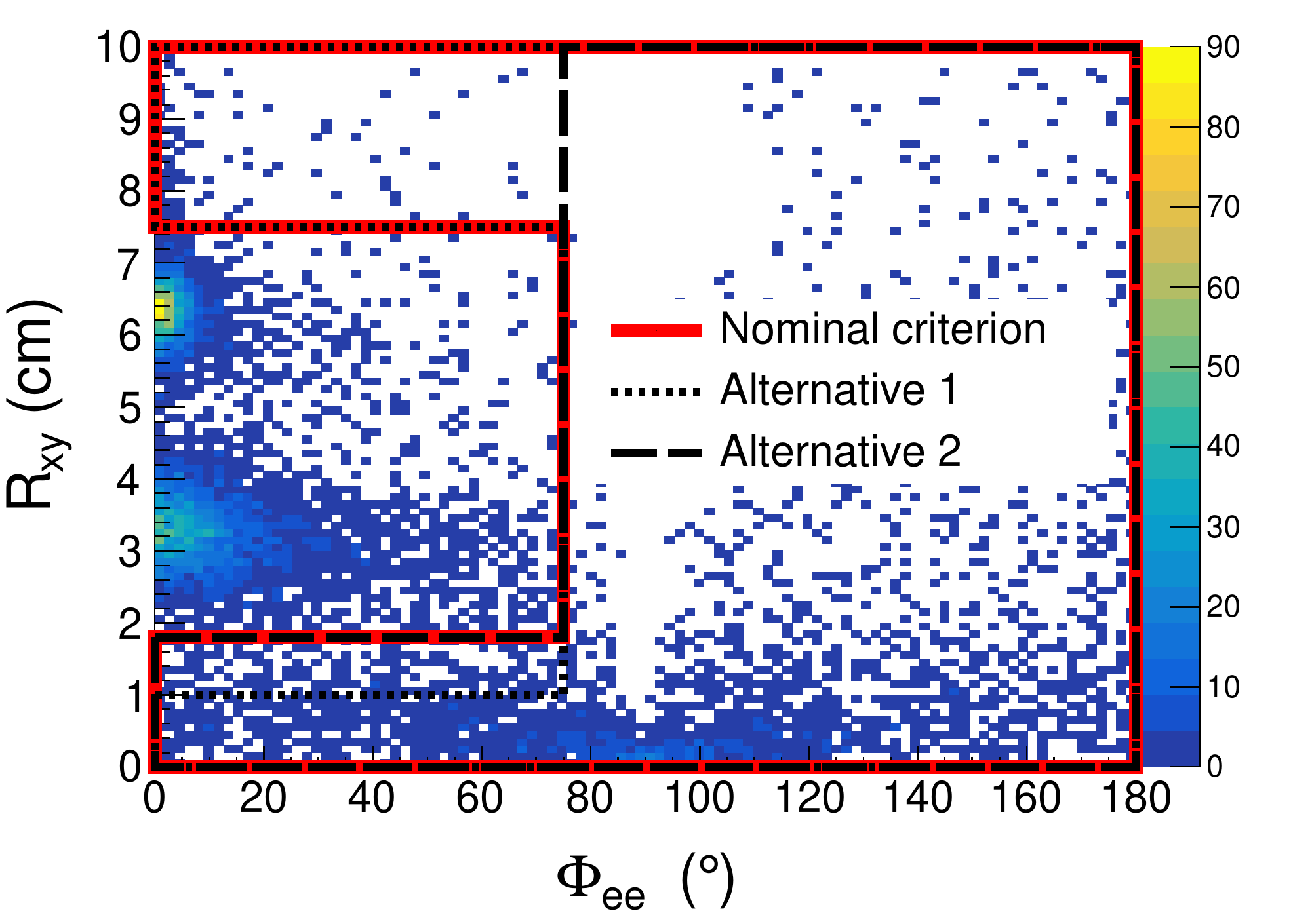}
\put(15,62){\color{red}\bf\large(b)}
\end{overpic}
\caption{Photon conversion veto criterion on the $M_{ee}^{BP}$ vs. $R_{xy}$ (a) and $\Phi_{ee}$ vs. $R_{xy}$ (b) plots from data. All events outside the solid (red) polygons are rejected. The dashed and dotted lines in each plot show the two alternative areas used to evaluate the systematic uncertainty due to the photon conversion veto.}
\label{fig:PhieeCut}
\end{figure}

After this preselection, an $\eta'$ peak is clearly visible in the invariant mass spectrum of $\pi^+\pi^-e^+e^-$ as shown in Fig.~\ref{fig:MeeMCData} (a). However, studies of an inclusive MC sample of $1.2 \times 10^9 \, J/\psi$ events show that there is significant background from $\eta' \to \pi^+\pi^-\gamma$ events, where the photon converts to an $e^+e^-$-pair in the beam pipe or the inner wall of the drift chamber. One would expect the invariant mass of such conversion pairs to be close to zero. However, the \mbox{BESIII} tracking algorithm uses the origin as a reference point for all tracks. This means that the direction of tracks that originate elsewhere will be mis-reconstructed. Hence, conversion pairs gain an artificial opening angle, and their reconstructed invariant masses are larger than the true values. Therefore, the conversion background appears as the large peak at about 0.015 GeV/$c^2$ in the $e^+e^-$ invariant mass distribution in Fig.~\ref{fig:MeeMCData} (b).
\begin{figure}
\centering
\includegraphics[scale=0.4]{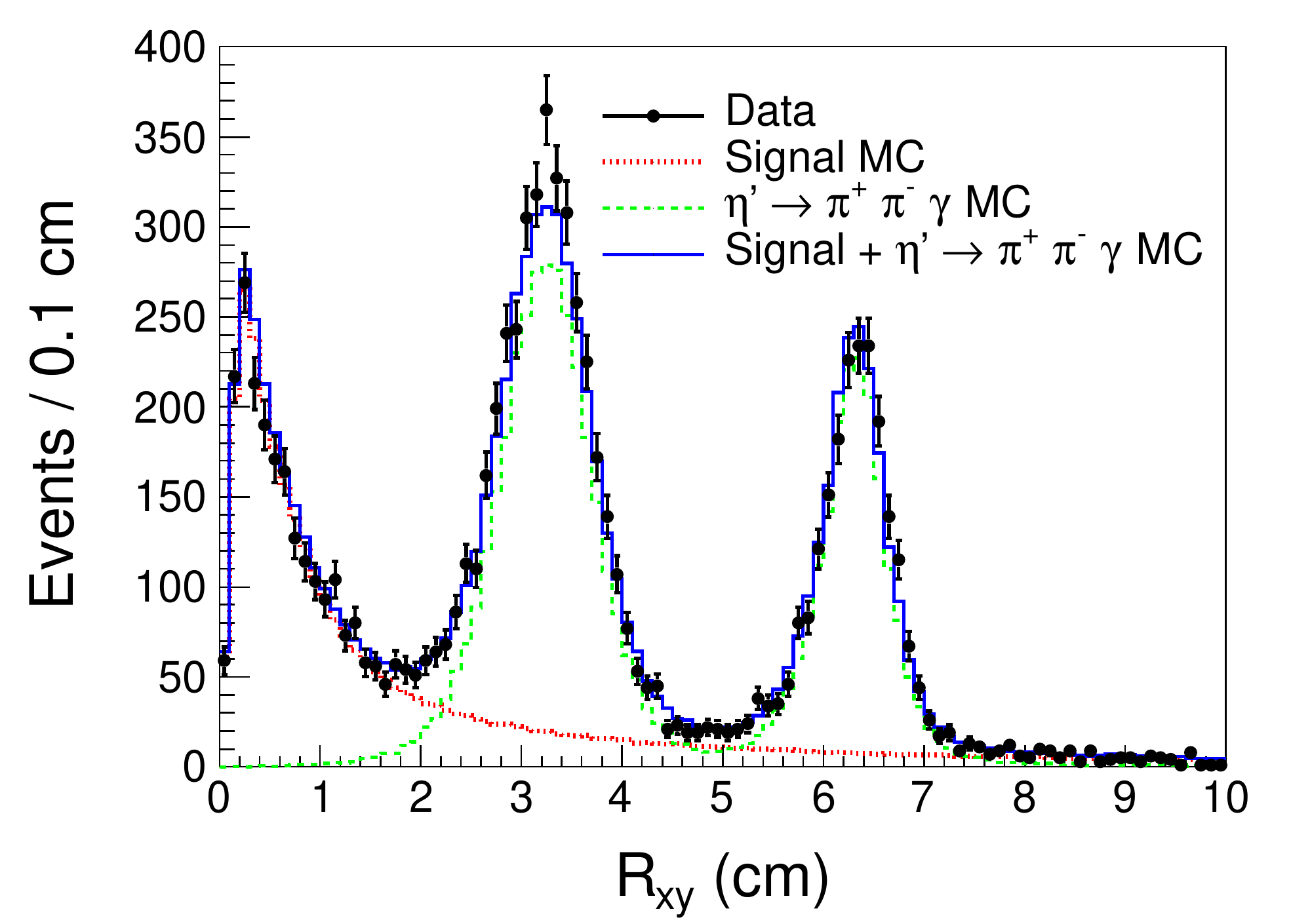}
\caption{Distribution of the distance of the $e^+e^-$-vertex from the IP in the transverse direction.  The dots with error bars correspond to data, the dotted (red) histogram represents the signal MC sample,  the dashed (green) histogram represents the $\eta' \to \pi^+ \pi^- \gamma$ MC sample, and the solid (blue) histogram is the sum of the two contributions. The two contributions are normalized according to their respective branching fractions and their efficiencies as determined from the MC simulation.}
\label{fig:Conv1D}
\end{figure}
By finding the intersection of the $e^+$- and $e^-$-helices in the $r-\phi$ projection, one acquires a measure of the distance from the $e^+e^-$-vertex position to the IP, $R_{xy}$. Figure~\ref{fig:Conv1D} shows $R_{xy}$ for the selected events. From this distribution, it is clear that the $e^+e^-$-pairs are created at three characteristic locations in the detector. The signal pairs effectively originate from the IP, whereas the conversion background pairs come from two different regions: at $R_{xy}\approx 3\;\textrm{cm}$, corresponding to the beam pipe, and at $R_{xy}\approx 6\;\textrm{cm}$, corresponding to the inner wall of the MDC. In order to reject photon conversion events and improve the signal-to-background ratio, two additional discriminating variables are introduced.
The first one, the invariant mass of the $e^+e^-$-pair at the beam pipe, $M_{ee}^{BP}$, is determined by changing the reference point of the helices of all $e^\pm$ tracks to their respective points of intersection with the beam pipe, and subsequently recalculating the momentum vectors. This procedure changes the direction of the vectors, but not their magnitudes.  For $e^+e^-$ pairs that originate from the IP, the opening angle is increased, and as a consequence, their invariant masses become larger than the true value. The momenta of the $e^+e^-$ pairs that were created in the beam pipe will instead be approximately parallel and the invariant mass close to the minimum value, $2m_e$. 
The second variable is the $z$-projection of the opening angle of the $e^+e^-$-pair, where $z$ is the  magnetic field direction, $\Phi_{ee}$~\cite{Adare:2009qk}. In conversion events, $\Phi_{ee}$ is expected to be close to zero, whereas in $\eta'$ decays it varies widely.  
In the two-dimensional distributions of $M_{ee}^{BP}$ vs. $R_{xy}$, and $\Phi_{ee}$ vs. $R_{xy}$, shown in  Fig.~\ref{fig:PhieeCut}, signal events and photon conversion events are well separated.  
By selecting appropriate regions of these distributions, the photon conversion events can be vetoed.  
First, we require $\Phi_{ee} < 75^\circ$ when 1.8 cm $ < R_{xy} <$ 7.5 cm.  
Then, in the $M_{ee}^{BP}$ vs. $R_{xy}$ distribution, we select all events to the high mass side of a curve defined by straight line segments between the points (0.004 GeV/$c^2$, 0 cm), (0.004 GeV/$c^2$, 2 cm), (0.03 GeV/$c^2$, 3 cm), and (0.07 GeV/$c^2$, 10 cm).  

\begin{figure}
\begin{overpic}[width=0.4\textwidth]{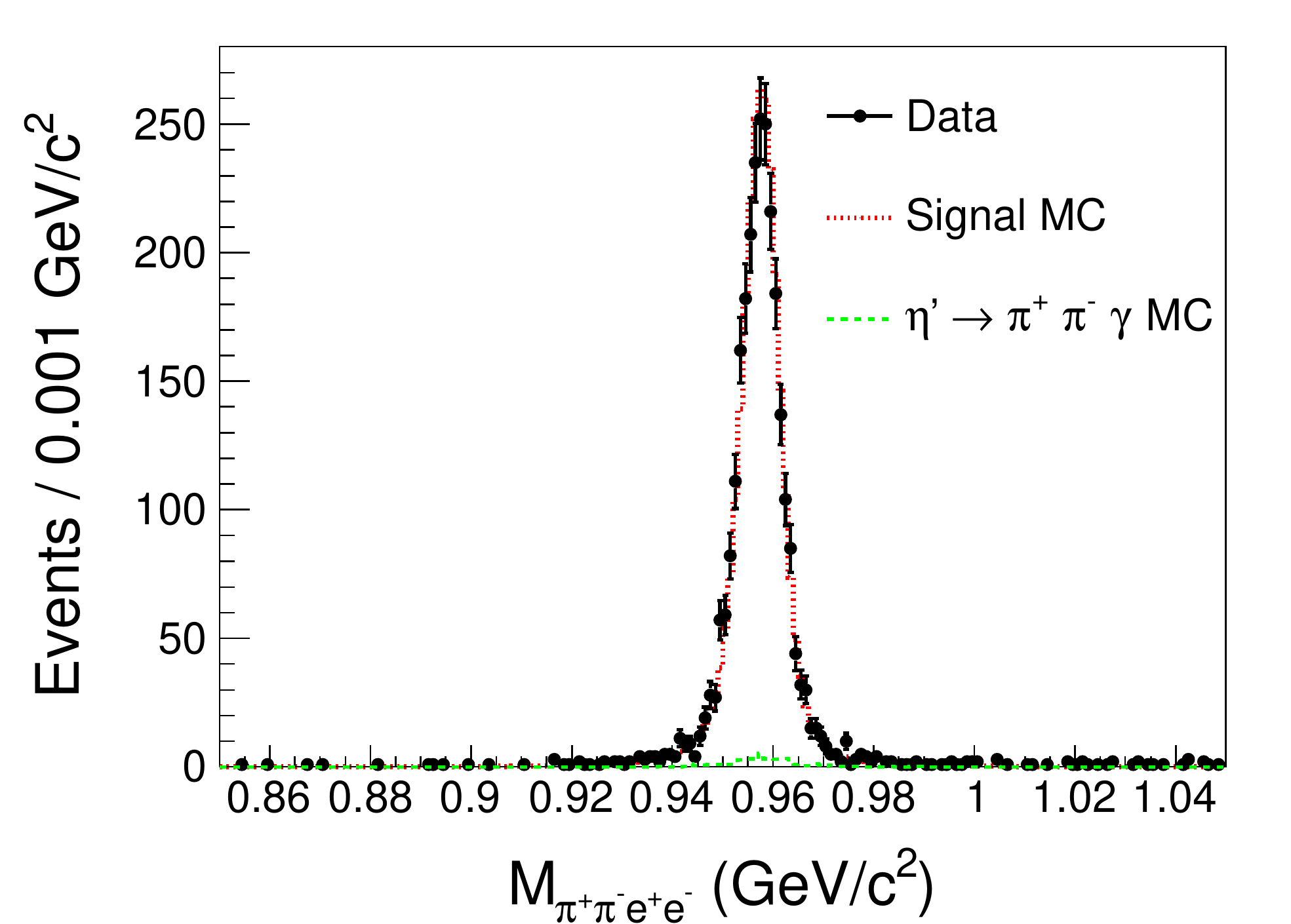}
\put(86,62){\color{red}\bf\large(a)}
\end{overpic}
\begin{overpic}[width=0.4\textwidth]{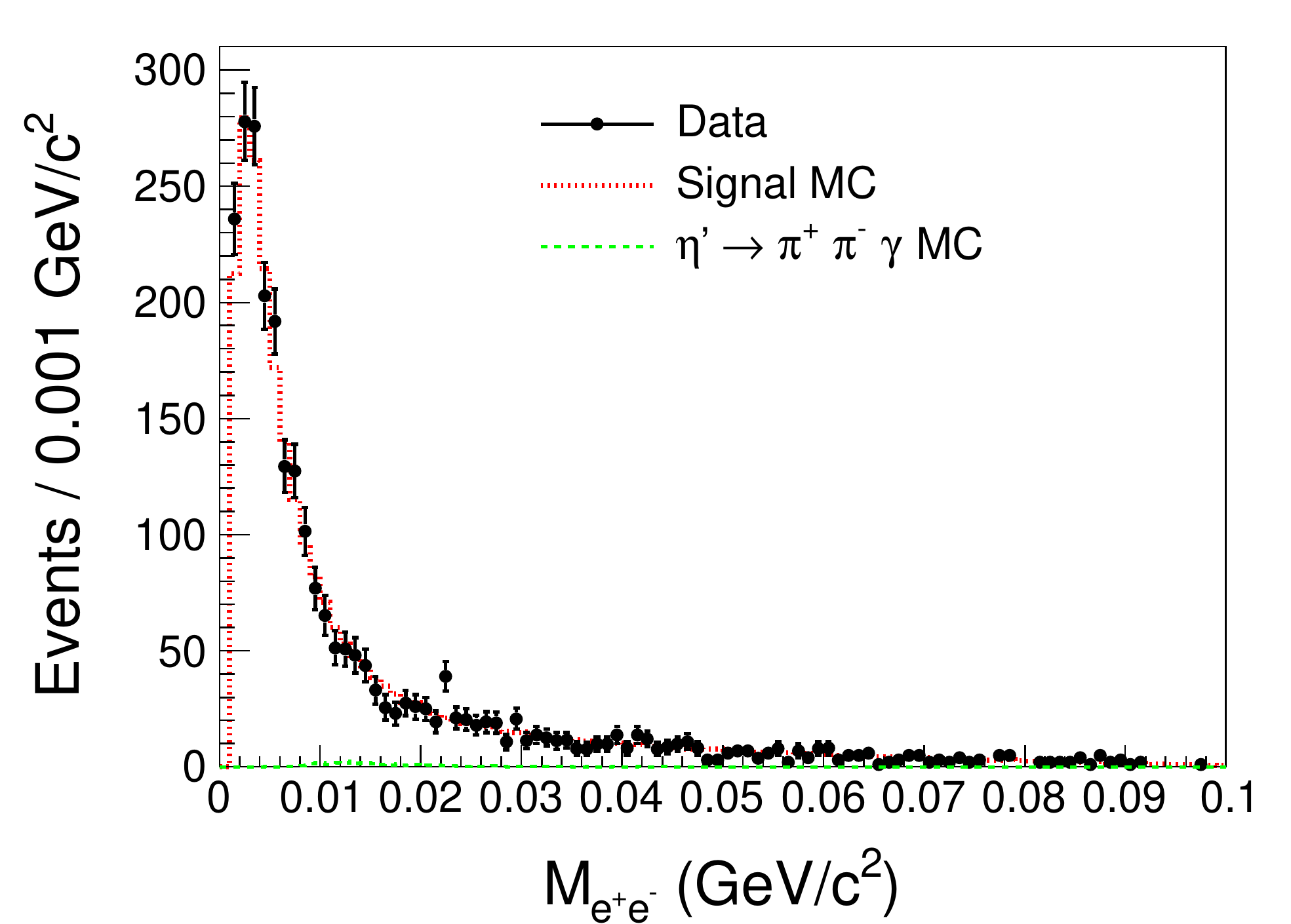}
\put(86,62){\color{red}\bf\large(b)}
\end{overpic}
\begin{overpic}[width=0.4\textwidth]{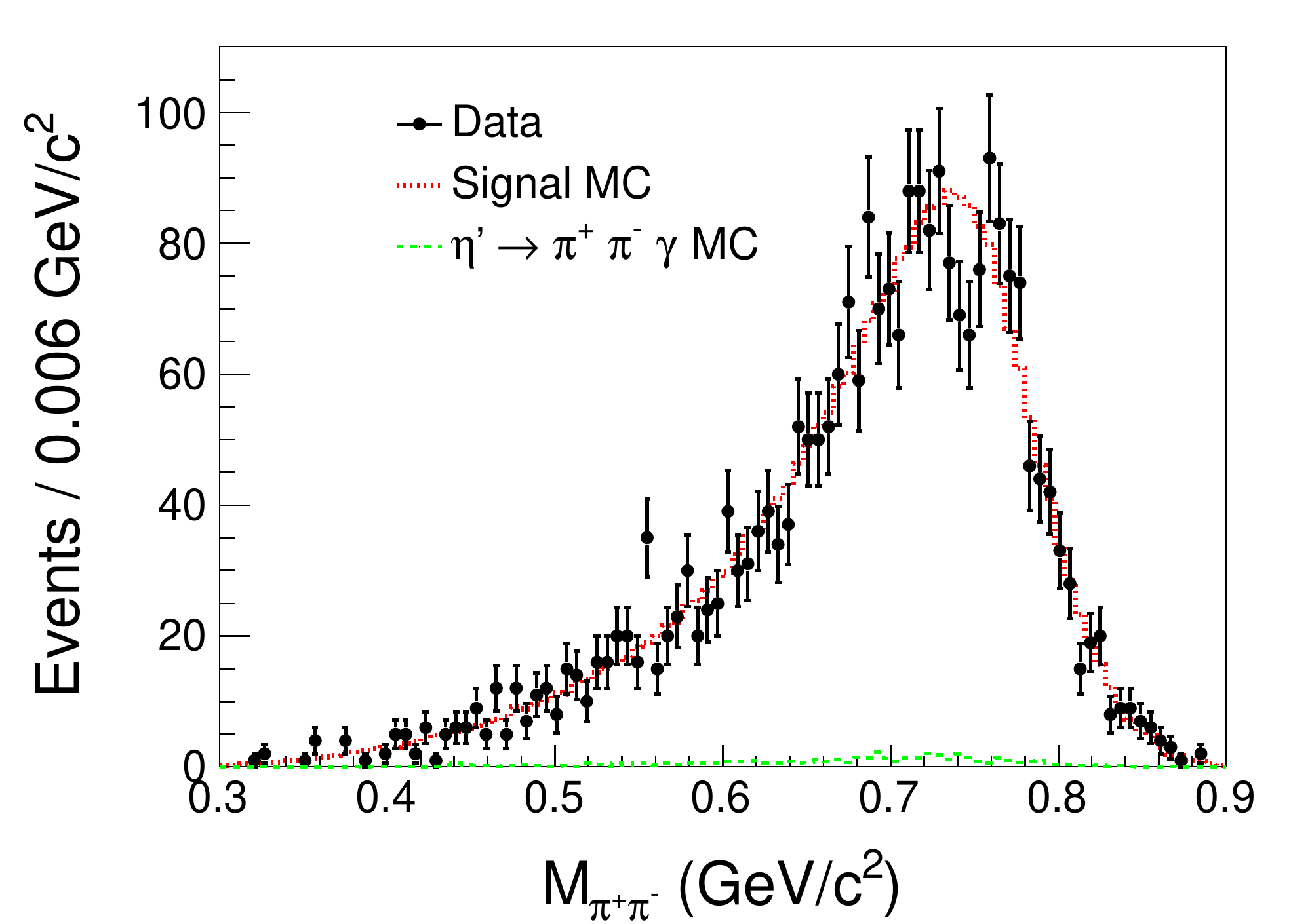}
\put(86,62){\color{red}\bf\large(c)}
\end{overpic}
    \caption{Invariant mass distributions of $\pi^+\pi^-e^+e^-$ (a), $e^+e^-$ (b) and $\pi^+\pi^-$ (c) for events after the photon conversion veto. Dots with error bars represent data, the dotted (red) histogram represents the signal MC sample, and the dashed (green) histogram the remaining background contribution from $\eta'\to\pi^+\pi^-\gamma$ events.}
    \label{fig:CutsMee}
\end{figure}

After the application of the photon conversion veto, most conversion events have been rejected, and the $\pi^+\pi^-e^+e^-$ mass peak is very clean, as shown in Fig.~\ref{fig:CutsMee}. Furthermore data and signal MC simulations are in good agreement in both the $e^+e^-$ and $\pi^+ \pi^-$ invariant mass distributions. 
Finally, we define the signal region as $\left| M_{\pi^+\pi^-e^+e^-} - m_{\eta'} \right|  < 0.02 \, \text{GeV/}c^2 $, where $m_{\eta'}$ is the mass of the $\eta'$ given by the PDG~\cite{PDG2020}. The signal purity of the sample is estimated to be 98\% based on our MC simulations of $\eta' \to \pi^+ \pi^- \gamma$. Accordingly, a residual contribution of 48 $\eta' \to \pi^+ \pi^- \gamma$ events is subtracted.  The number of combinatorial background events remaining is estimated from the $\eta'$-sidebands in the $M_{\pi^+\pi^-e^+e^-}$ distribution, defined as $0.06 \, \text{GeV/}c^2 < | M_{\pi^+\pi^-e^+e^-} - m_{\eta'}| < 0.08 \, \text{GeV/}c^2 $. The number of events remaining after these subtractions is taken as the final yield for $\eta' \to \pi^+\pi^- e^+e^-$ and is listed in Table~\ref{tab:EffN} together with the detection efficiency.

\begin{table}[t]
\centering
\caption{Efficiencies and event yields for the signal and normalization channels}
\label{tab:EffN}
\begin{tabular}{l|l|l}
\hline \hline 
Channel & $\varepsilon [\%]$ & Yield \\ \hline 
$\eta' \to \pi^+\pi^-e^+e^-$ & $15.25 \pm 0.01$ & $2584 \pm 52 $ \\
$\eta' \to \pi^+\pi^-\gamma$ & $38.09 \pm 0.01$ & $786200 \pm 900$ \\ \hline \hline
\end{tabular}
\end{table}

\begin{figure}
\centering
\includegraphics[scale=0.4]{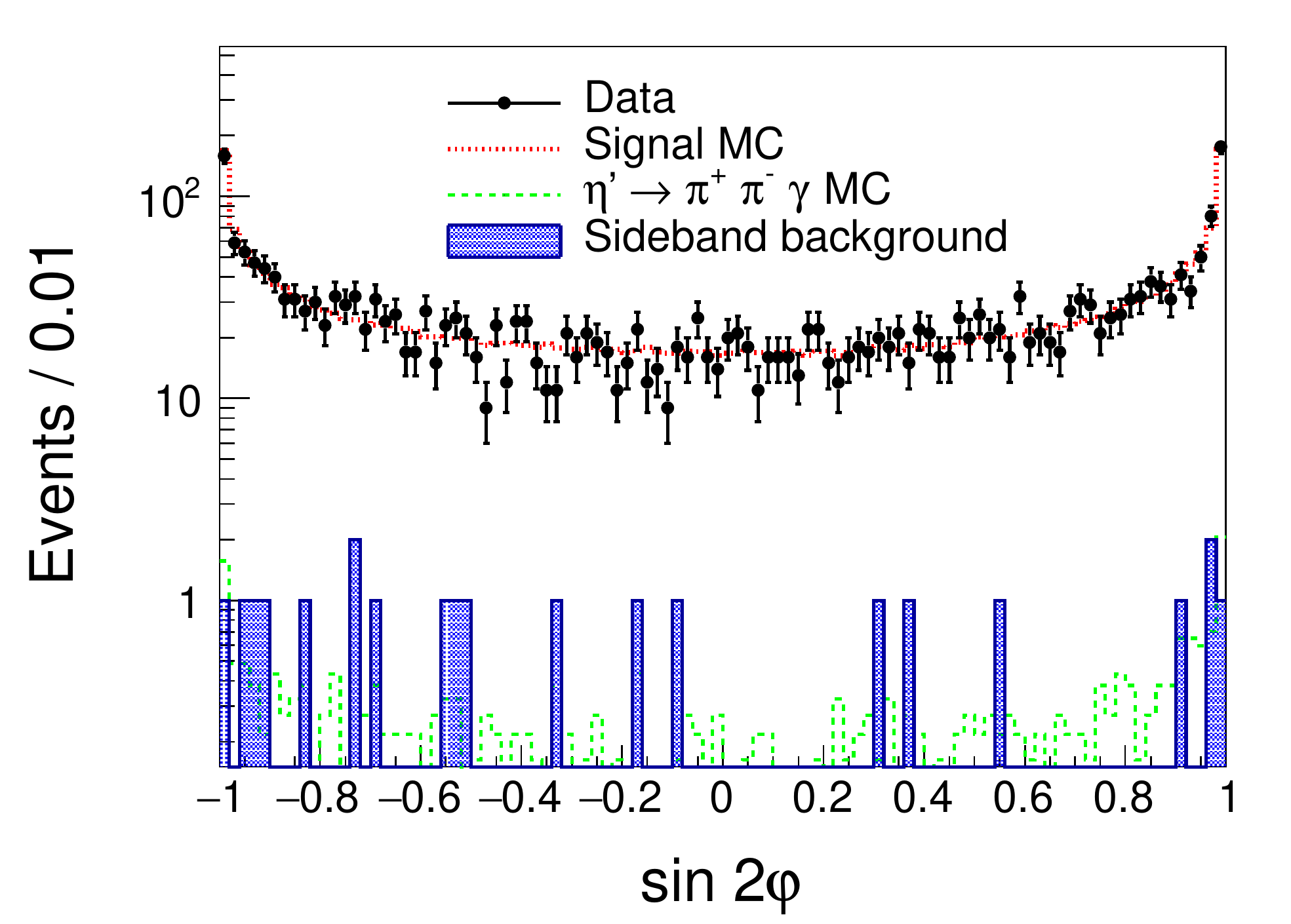}
\caption{Distribution of $\sintwophi$ after the photon conversion veto. Dots with error bars represent data, the dotted (red) histogram represents the signal MC sample generated with a symmetric $\sintwophi$ distribution, the dashed (green) histogram represents the remaining background contribution from $\eta' \to \pi^+ \pi^- \gamma$ events and the filled (blue) histogram the $\eta'$ sideband contribution.}
\label{fig:SinCos}
\end{figure}

The quantity $\sintwophi$ is calculated for each event using the method described in Ref.~\cite{Ambrosino2009283}. The orientation of the decay planes is given by the unit normal vectors defined as:
\begin{align}
\hat{n}_e& = \frac{{\bf p}_{e^+} \cross {\bf p}_{e^-}}{\left| {\bf p}_{e^+} \cross {\bf p}_{e^-}\right|} \ \text{and}\
\hat{n}_\pi = \frac{{\bf p}_{\pi^+} \cross {\bf p}_{\pi^-}}{\left| {\bf p}_{\pi^+} \cross{\bf p}_{\pi^-}\right|} \ ,
\label{eq:Sin2phiBegin}
\end{align}
where ${\bf p}_{\pi^\pm}$, ${\bf p}_{e^\pm}$ are the momenta of the pions and electrons in the $\eta'$ rest frame, respectively. It follows from the properties of the dot product that:
\begin{equation}
\hat{n}_e \cdot \hat{n}_\pi = \cos \varphi  \ \text{and}\ \left| \hat{n}_e \cross \hat{n}_\pi \right|=\left| \sin \varphi \right| \ ,
\end{equation}
where $\varphi$ is the angle between the normal vectors, i.e., the asymmetry angle. 
The sign of $\sin \varphi$ is obtained by using the direction of the unit vector along the intersection of the two planes (see Fig.~\ref{fig:AsymmIll}) 
\begin{equation}
\hat{z}  = \frac{{\bf p}_{e^+} + {\bf p}_{e^-}}{\left| {\bf p}_{e^+} + {\bf p}_{e^-} \right| } 
\end{equation}
and then $\sin \varphi$ is calculated as
\begin{equation}
\sin \varphi = (\hat{n}_e \cross \hat{n}_\pi) \cdot \hat{z} \ .
\end{equation}
The final expression for $\sintwophi$ is
\begin{equation}
\sintwophi = 2 \sin \varphi \cos \varphi = 2 \left[(\hat{n}_e \cross \hat{n}_\pi) \cdot \hat{z}\right] (\hat{n}_e \cdot \hat{n}_\pi) \ .
\label{eq:Sin2phiEnd}
\end{equation}
%
The $\sintwophi$ distributions from data and the $\eta' \to \pi^+ \pi^- e^+ e^-$ MC sample are shown in Fig.~\ref{fig:SinCos}. The signal yields for  events with $\sintwophi > 0$ and $\sintwophi < 0$ are determined by fits to the two $ \pi^+ \pi^- e^+e^-$ invariant mass distributions. The fits are performed in the range
$0.91 \text{ GeV/}c^2 < M_{\pi^+\pi^-e^+e^-} < 1.00 \text{ GeV/}c^2$,
where the signal is represented by the MC lineshape convolved with a Gaussian function to account for differences in detector resolution between data and MC simulations. Combinatorial background is represented by a second order Chebychev polynomial. Peaking backgrounds are subtracted based on MC simulations.

Limited momentum resolution will cause some fraction $\alpha$ of events with a true value $\sintwophi<0$ to be reconstructed with a value $\sintwophi>0$ and vice versa. This migration effect must be corrected for in order to obtain an unbiased estimate of the asymmetry of the sample. By studying our signal MC sample, we find that the fraction of events that migrate from $\sintwophi>0$ to $\sintwophi<0$ is the same as that which migrates from $\sintwophi<0$ to $\sintwophi>0$, and we estimate it to be $21.3 \pm0.1\%$. Given that the efficiencies in the two regions are the same within the uncertainty, see Table~\ref{tab:RegYields}, the asymmetry corrected for bin migration in $\sintwophi$ is
\begin{equation}
    \mathcal{A}_{\varphi, \, \text{corr.}} = \frac{\mathcal{A}_{\varphi,\, \text{rec.}}}{1-2\alpha},
\end{equation}
where $\mathcal{A}_{\varphi,\, \text{rec.}}$ is the asymmetry acquired by inserting the event yields into Eq.~\ref{eq:AsymmExp}.

\begin{figure}
    \centering
    \includegraphics[scale=0.4]{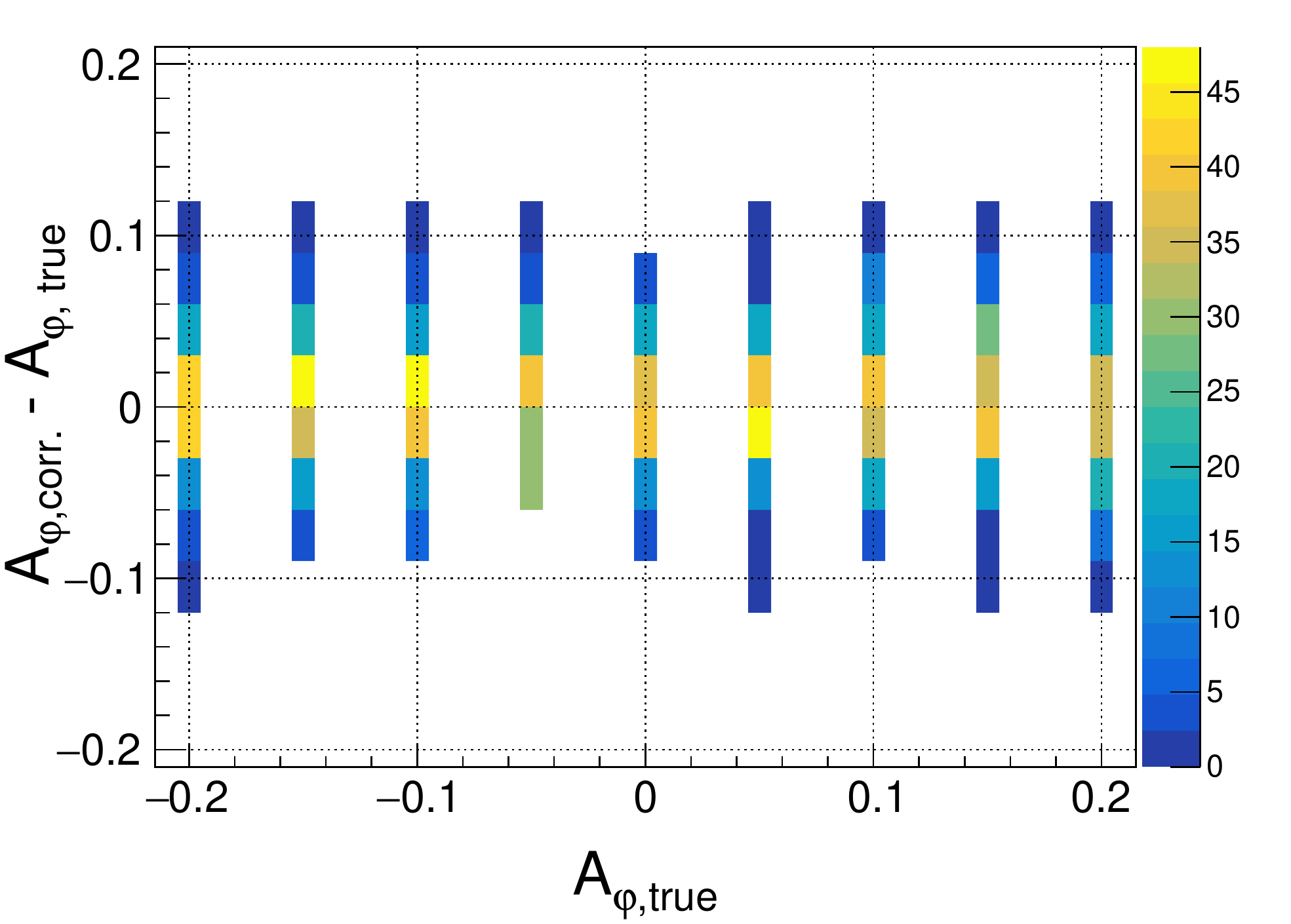}
    \caption{Difference between the corrected reconstructed asymmetry and the true asymmetry in nine MC ensembles. The ensembles are generated with asymmetries ranging from -20 \% to +20 \% in steps of 5 \% and each consists of 125 samples that match our data in size. The bin width along the $y$-direction is roughly equal to the root-mean-square deviation.}
    \label{fig:TrueRecAsymm}
\end{figure}

The accuracy of the method used to estimate the asymmetry is evaluated by applying it to nine MC ensembles generated with asymmetries ranging from $-$20\% to +20\% in steps of 5\%. In each ensemble there are 125 samples, and each sample matches our data in size. The asymmetry is extracted from each sample and compared to the expected value as is shown in Fig.~\ref{fig:TrueRecAsymm}. It is clear that, on average, the true value of the asymmetry can be reliably reproduced.

\begin{table}[t]
\centering
\caption{Efficiencies and event yields in the two regions of the decay plane angle $\varphi $. }
\label{tab:RegYields}
\begin{tabular}{l|l|l}
\hline \hline 
Region & $\varepsilon [\%]$ & Yield \\ \hline 
$\sintwophi > 0$ & $15.95 \pm 0.02$  & $1331 \pm 40$ \\
$\sintwophi < 0$ & $15.93 \pm 0.02 $ & $1287 \pm 37$ \\ \hline \hline
\end{tabular}
\end{table}

\subsection{Normalization Channel: $\eta' \to \pi^+ \pi^- \gamma$}
 To select the decay $\eta' \to \pi^+\pi^-\gamma$, at least two photons and exactly two charged tracks with zero net charge are required. The radiative photon coming from the initial $J/\psi$ decay is monoenergetic at $E_{\gamma_{\textrm{rad}}}=1.4$ GeV and we require one photon to have an energy in a $20$ MeV window around that value. In all other aspects, the event selection procedure for this channel is identical to that described in the previous section. The optimal cutoff for the kinematic fit and PID is found to be $\chi^2_{4C+PID} < 140$. After these initial selection criteria, the inclusive MC sample shows significant contamination from events with $\pi^0\to \gamma \gamma$ or $\eta \to \gamma \gamma$. Such events can be separated from the signal by studying the energy distribution of the photon from the $\eta'$ decay, and requiring $E_\gamma > 0.15$ GeV.
\begin{figure}
\centering
\includegraphics[scale=0.4]{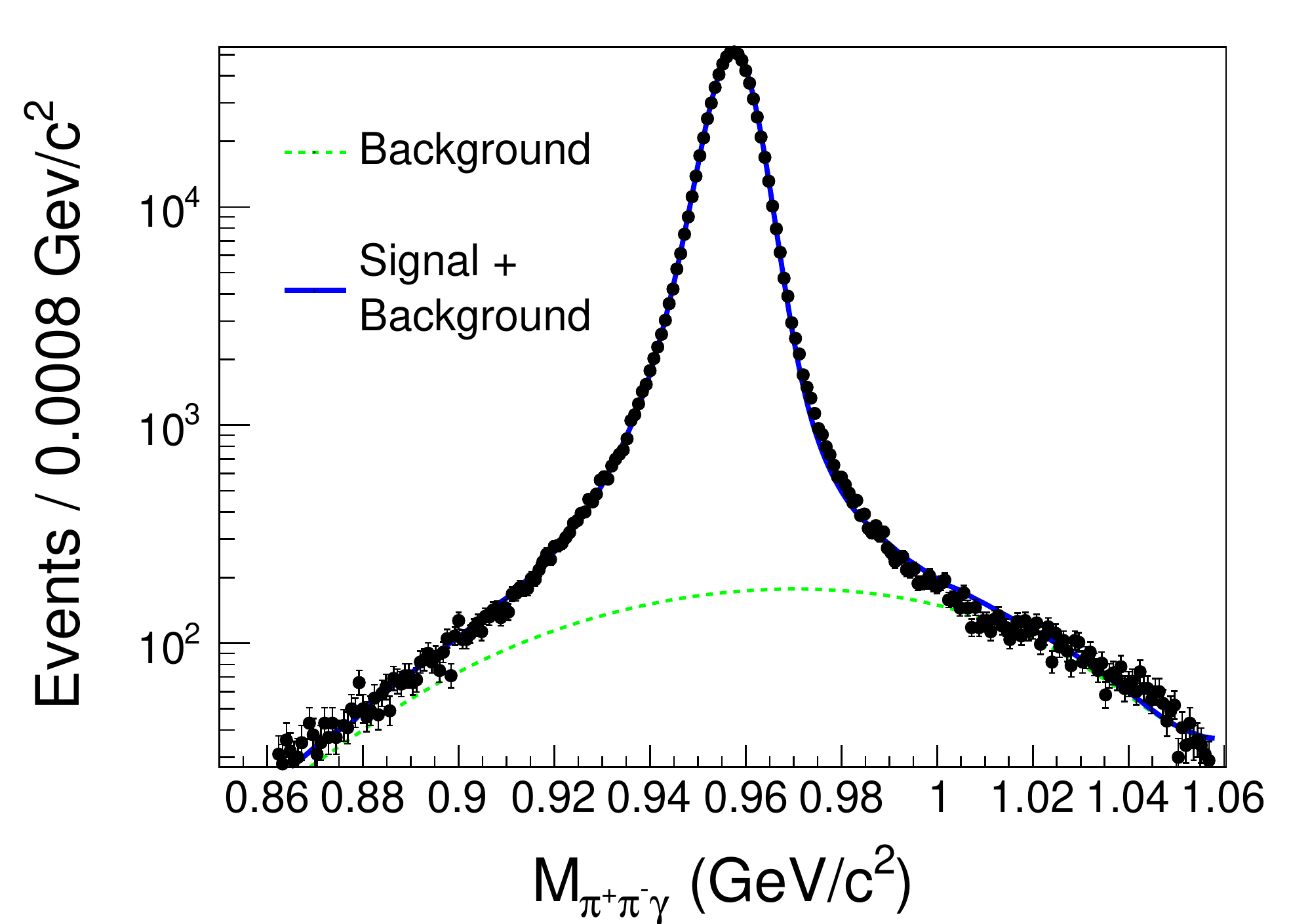}
\caption{Fit to the invariant mass distribution of $\pi^+ \pi^- \gamma$. Dots with error bars represent data, the dashed (green) line represents continuous background, and the solid (blue) line is the sum of the signal and background contributions.}
\label{fig:EtapFit}
\end{figure}
The final signal yield is extracted by a fit to the $\pi^+\pi^-\gamma$-invariant mass distribution shown in Fig.~\ref{fig:EtapFit}. The signal is represented by the MC line shape convolved with a Gaussian function in order to take into account the difference in resolution between data and MC simulations. Continuous background is represented by a second order Chebychev polynomial. The yield and efficiency for the normalization channel are listed in Table~\ref{tab:EffN}.

\section{Systematic Uncertainties}
A list of all sources of systematic uncertainties is given in Table~\ref{tab:SystSum}
and details of how their contributions are estimated are given below:
\begin{itemize}
\item  {\bf MDC Tracking } The data-MC efficiency difference for pion track-finding has been studied using a control sample of $J/\psi \to p \bar{p} \pi^+ \pi^-$ events. For the tracking of electrons, a mixed sample of  $e^+e^- \to \gamma e^+e^-$ at the $J/\psi$ meson mass and $J/\psi \to e^+e^- \left( \gamma_{FSR} \right)$ was used. In both cases, the data-MC difference, $\Delta_{syst.}$, is extracted as a function of the particle momentum and the cosine of the polar angle. Subsequently, each event in the MC samples is re-weighted by a factor $(1 + \Delta_{syst.})$. The branching fraction and asymmetry parameter are recalculated with efficiencies determined from the re-weighted MC sample, and the difference from the original result is taken as the systematic uncertainty. 
\item {\bf PID} The effect of the difference in PID efficiency between data and MC is evaluated in the same way as for the MDC tracking efficiency above. The control samples used are $J/\psi \to \pi^+ \pi^- \pi^0$ for pions and  $e^+e^- \to \gamma e^+e^-$ at the $J/\psi$ meson mass and $J/\psi \to e^+e^- \left( \gamma_{FSR} \right)$ for electrons. 
\item {\bf Photon Reconstruction} The data-MC difference in the photon reconstruction efficiency has been studied based on a number of control samples, including $J/\psi \to \rho^0 \pi^0$, $e^+e^- \to \gamma \gamma$~\cite{Ablikim:2010zn}, and $e^+e^- \to \gamma \mu^+ \mu^-$. It has been found that the average difference is $0.5 \%$ in the EMC barrel ( $|\cos \theta_\gamma| < 0.80$), and $1.5 \% $ in the endcaps ($ 0.86 \leq|\cos \theta_\gamma| \leq 0.92$). As for MDC tracking and PID, the MC samples are re-weighted event-by-event to correct these shifts.
\item {\bf 4C Kinematic Fit} Using a control sample of $J/\psi \to \phi f_0(980)$, a set of correction factors for the track helix parameters is used within the \mbox{BESIII} collaboration to improve the agreement of the 4C kinematic fit performance between MC and data. The difference in branching fraction and asymmetry with and without these corrections is taken as the systematic uncertainty due to the kinematic fit. 
\item {\bf\boldmath $\eta'$ Mass Window} The effect of differences in the $\eta'$ mass resolution between MC and data are evaluated by smearing the $\pi^+ \pi^- e^+e^-$ and $\pi^+\pi^- \gamma$ invariant mass distributions according to the difference in resolution between data and MC as extracted from the fit in Fig.~\ref{fig:EtapFit}. The difference between the original result and the result with smearing is taken as the systematic uncertainty. 
\item {\bf Photon Conversion Veto} Systematic effects from the photon conversion veto are evaluated by introducing two alternative criteria for the invariant mass at beam pipe, and two for the opening angle $\Phi_{ee}$, Fig.~\ref{fig:PhieeCut}. The branching fraction and asymmetry parameters are calculated for all combinations of the original and alternative selection criteria. Half the largest difference from the original result is taken as the systematic uncertainty. 
\item {\bf Normalization} This is the combined systematic uncertainty for the normalization channel, including both the uncertainty on the known branching fraction $ {\cal B}(\eta' \to \pi^+ \pi^- \gamma)$ and uncertainties due to tracking, photon reconstruction, the requirement $E_{\gamma_{\eta'}}>0.15$~GeV/$c$ and effects from the range and background shape used in the fit to the $\pi^+\pi^-\gamma$ invariant mass distribution in Fig.~\ref{fig:EtapFit}. The photon energy cut, the fit range, and the background polynomial order are each varied and the largest changes caused by each are taken as systematic uncertainties. The value for $ {\cal B}(\eta' \to \pi^+ \pi^- \gamma)$ listed in the PDG is solely based on Ref.~\cite{bes3normchan}, using the same dataset. Thus, we take into account the strong correlation between the literature value of $ {\cal B}(\eta' \to \pi^+ \pi^- \gamma)$ and our analysis of the normalization channel. Due to small differences in the selection procedure, our final sample is a subset of the sample used in Ref.~\cite{bes3normchan} with a correlation coefficient of 0.93 for the statistical uncertainty. For systematic uncertainties, we assume correlation coefficients to be one in case of sources that appear in both works, and zero for sources that appear only in one of the two works. The former case includes the systematic uncertainties regarding tracking, photon reconstruction, kinematic fitting, as well as range and background shape used in the fit.
This systematic uncertainty only applies to the branching fraction.
\end{itemize}
\begin{table}[t]
\caption{Systematic uncertainties on ${\cal B}(\eta' \rightarrow \pi^+ \pi^- e^+ e^-)$ and $\asymm$. The systematic effects of MDC tracking, PID, and photon reconstruction on $\asymm$ are considered negligible compared to the other contributions.}
\label{tab:SystSum}
\begin{tabular}{l|c|c}
\hline \hline 
Source & ${\cal B}(\eta' \rightarrow \pi^+ \pi^- e^+ e^-)$ & $\asymm$ $\left(\times10^{-2}\right)$\\
       &                        [\%]                             &                                \\
\hline 
MDC tracking & 0.7 & -\\
PID & 3.0 & -\\
Photon Reconstruction & 0.6 & -\\
4C Kinematic Fit & 0.3 & 0.5 \\
$\eta'$ mass window & 0.4  & - \\
Photon Conversion Veto & 0.8 & 0.9 \\
Normalization  & 1.3 & - \\ \hline
Total & 3.5 & 1.1\\ \hline \hline
\end{tabular}
\end{table}

\section{Results}

\noindent
Given the efficiencies and event yields in Table~\ref{tab:EffN}, the ratio ${\cal B}(\eta' \rightarrow \pi^ +\pi^- e^+ e^-) / {\cal B}(\eta' \rightarrow \pi^ +\pi^- \gamma)$ is determined to be $\left( 8.20\pm0.16_{stat.}\pm0.27_{syst.} \right) \times 10^{-3}$. 
Taking into account the world average of the branching fraction of $\eta' \rightarrow \pi^+ \pi^- \gamma$~\cite{PDG2020}, the branching fraction of $\eta' \to \pi^+ \pi^- e^+ e^-$ is $(2.42\pm0.05_{stat.}\pm0.08_{syst.}) \times 10^{-3}$. The statistical uncertainty has been improved by a factor of two compared to the last \mbox{BESIII} result, which is superseded by the result of this work due to the inclusion of the additional data collected in 2012.
Ref.~\cite{Petri:2010ea} predicts the branching fraction from two different VMD models (the hidden gauge model and the modified VMD model), to be $(2.17\pm 0.21) \times 10^{-3}$ and $(2.27\pm 0.13) \times 10^{-3}$ respectively. The unitary chiral perturbation theory approach of Ref.~\cite{Borasoy:2007dw} yields a branching fraction of $(2.13^{+0.17}_{-0.31})\times 10^{-3}$. Our result is consistent with all three predictions; it is about one standard deviation higher than each of them. The CP-violating asymmetry is determined to be
$(2.9\pm3.7\pm1.1)\%$, which is consistent with zero. This work achieves precision comparable to that of the asymmetry measurement in the $K^0_L\to\pi^+\pi^-e^+e^-$ decay~\cite{PhysRevLett.84.408, IconomidouFayard:2001ww}, but the size of the asymmetry determined here is significantly smaller than the SM driven effect of $(14\pm2)\%$ observed in the $K^0_L$ decay.

\section{Acknowledgement}
The BESIII collaboration thanks the staff of BEPCII and the IHEP computing center for their strong support. This work is supported in part by National Key Basic Research Program of China under Contract No. 2015CB856700; National Natural Science Foundation of China (NSFC) under Contracts Nos. 11625523, 11635010, 11735014, 11675184; National Natural Science Foundation of China (NSFC) under Contract No. 11835012; the Chinese Academy of Sciences (CAS) Large-Scale Scientific Facility Program; Joint Large-Scale Scientific Facility Funds of the NSFC and CAS under Contracts Nos. U1532257, U1532258, U1732263, U1832207; CAS Key Research Program of Frontier Sciences under Contracts Nos. QYZDJ-SSW-SLH003, QYZDJ-SSW-SLH040; 100 Talents Program of CAS; INPAC and Shanghai Key Laboratory for Particle Physics and Cosmology; German Research Foundation DFG under Contract No. Collaborative Research Center CRC 1044; Istituto Nazionale di Fisica Nucleare, Italy; Koninklijke Nederlandse Akademie van Wetenschappen (KNAW) under Contract No. 530-4CDP03; Ministry of Development of Turkey under Contract No. DPT2006K-120470; National Science and Technology fund; The Knut and Alice Wallenberg Foundation (Sweden) under Contract No. 2016.0157; The Royal Society, UK under Contract No. DH160214; The Swedish Research Council; U. S. Department of Energy under Contracts Nos. DE-FG02-05ER41374, DE-SC-0010118, DE-SC-0012069; University of Groningen (RuG); The Helmholtzzentrum fuer Schwerionenforschung GmbH (GSI), Darmstadt and the Olle Engkvist Foundation (Sweden) under Contract No. 200-0605.

\bibliographystyle{h-physrev5_mod}

\end{document}